\documentclass[twocolumn,english,aps,pre,showpacs,floats]{revtex4}
\usepackage[T1]{fontenc}
\usepackage[latin1]{inputenc}
\usepackage{color}
\usepackage{float}
\usepackage{bm}
\usepackage{amsmath}
\usepackage{amssymb}
\usepackage{graphicx}

\makeatletter
\@ifundefined{textcolor}{}
{%
 \definecolor{BLACK}{gray}{0}
 \definecolor{WHITE}{gray}{1}
 \definecolor{RED}{rgb}{1,0,0}
 \definecolor{GREEN}{rgb}{0,1,0}
 \definecolor{BLUE}{rgb}{0,0,1}
 \definecolor{CYAN}{cmyk}{1,0,0,0}
 \definecolor{MAGENTA}{cmyk}{0,1,0,0}
 \definecolor{YELLOW}{cmyk}{0,0,1,0}
}

\usepackage{dcolumn}

\usepackage{bm}
\usepackage{mathrsfs}

\DeclareFontFamily{U}{calligra}{}
\DeclareFontShape{U}{calligra}{m}{n}{<->callig15}{}

\providecommand{\second}{\prime\prime}

\usepackage{babel}

\makeatother

\usepackage{babel}
\begin{document}
\title{Time profile of temperature rise in assemblies of nanomagnets}
\author{J.-L. Déjardin and H. Kachkachi}
\email{hamid.kachkachi@univ-perp.fr}

\affiliation{Université de Perpignan via Domitia, Lab. PROMES CNRS UPR8521, Rambla
de la Thermodynamique, Tecnosud, 66100 Perpignan, France}
\begin{abstract}
We compute the heat generated by (non-interacting) nanomagnets subjected
to an alternating magnetic field (AMF) and study its transfer to the
hosting medium and environment. For the first task, we compute the
heat generated by the nanomagnets (or the specific absorption rate
$\mathfrak{S}$) using the ac susceptibility in the linear regime.
For the second task, the loss of heat to the environment is modeled
with the help of a balance (macroscopic) equation based on Newton's
law of cooling. This equation is solved both numerically and analytically
for a generic ferrofluid and the analytical solution renders a very
good approximation to the general balance equation. Then, we investigate
the effects of AMF frequency and amplitude on the temperature elevation
during its temporal evolution. Finally, using the available experimental
data for maghemite and magnetite ferrofluids, we discuss the behavior
of Newton's heat transfer coefficient in terms of the AMF amplitude
and frequency. These results could trigger experimental investigations
of this coefficient which characterizes the rate of heating in a ferrofluid,
with the aim to build more refined models for the mechanisms of heat
generation and its diffusion in ferrofluids used in magnetic hyperthermia. 
\end{abstract}
\date{\today}
\maketitle

\section{Introduction}

The process of magnetic hyperthermia (MH) occurring in assemblies
of magnetic nanoparticles, or nanomagnets (NM), offers a significant
advantage over the traditional methods for cancer treatment \citep{OvergaardEtAl_TF85,vanderZee_ao02,JohannsenEtAl_ijh05,SkumielEtAl_ijt,MuraseEtAl_physmed13}
based on the necrosis of the cancerous cells by irradiation \citep{StrefferEtAl_TF85},
as it avoids the side effects of the latter by a directed and localized
heating. Moreover, MH has been shown to kill cells faster than the
traditional methods and this reduces the therapy administration time
\citep{RodriguezEtAl_ijn11}. However, at the present time the mechanism
by which the heat is generated and diffused in the assembly of NM
is not fully understood.

In MH, the NM are heated with the help of an external AC magnetic
field, that is a time-dependent (alternating) magnetic field (AMF),
and the optimization of the whole process depends on several factors,
such as the magnetic field itself (amplitude and frequency), NM size
and concentration, and solution viscosity \citep{Carrey_JAP2011,Mehdaoui_AFM2011,martinez2013learning,condeetal15jpcc,koslap2015nanotechrev}.
The efficiency of MH may be assessed through the \emph{specific absorption
rate} (SAR), usually given in Watt per gram (W/g). The SAR has been
investigated by many research groups by computing the area of the
hysteresis loop of the magnetization as a function of the AC field
intensity, for a given frequency \citep{Lacroix_etal_JAP2009,Carrey_JAP2011,Mehdaoui_AFM2011,mehdaouietal12apl,martinez2013learning,condeetal15jpcc}.
With the desire to obtain (approximate) analytical expressions, an
alternative approach was adopted in Refs. \citep{dejardin_etal_SAR_JAP_2017,DejVerKach_jap20,DejVerKach_springer20}
which exploits the fact that the system absorption of the electromagnetic
energy brought in by the AMF may be described by the out-of-phase
component of the dynamic response function, the AC susceptibility.

Another observable that is being actively used and which is quite
relevant for characterizing MH and assessing its efficiency is the
temperature rise after the start of heating, upon varying the AMF
characteristics and the sample's properties \citep{muraseEtal_rpt11,AndreuNatividad_ijh13,SHAH201596}.
This technique is useful as it allows for a more precise control of
the temperature rise in MH in order to reduce the risk of heating
the surrounding healthy cells. A detailed discussion of this calorimetric
method and its comparison with other methods is presented in Ref.
\citep{AndreuNatividad_ijh13}. Various empirical equations are used
to model the temperature rise in MH and the initial slope of $T\left(t\right)$
is used for estimating the SAR on the ground of two assumptions: i)
upon switching on the AMF, the sample temperature remains homogeneous,
and ii) the heat losses are negligible at short times. However, this
initial-slope technique has a few drawbacks exemplified, for instance,
by the fact that it is based on a simplified model which ignores the
heat diffusion across the sample and that the determination of the
initial-time slope is performed at a transient state of the sample
temperature \citep{AndreuNatividad_ijh13,AfrandEtAl_ichmt16,EsmaeiliEtAl_ate17,ShaEtAl_ate17}.
It also ignores the heat flow from the sample into its environment.
It would then be desirable to go a step further and take into consideration
both the heating of the sample via the excitation of NM by the AMF
and the diffusion of the thus-generated heat into the sample and ultimately
the heat loss toward the environment. However, the issue of heat flow
in suspensions is rather involved and one has to proceed in steps
to separately investigate each one of these mechanisms. Regarding
the efficiency of heat flow in nanofluids and, in general heat transport
at the nanoscale, there is a rich literature studying the underlying
mechanisms with attempts to extend Fourier's law, see for instance
Refs. \citep{Keblinski_IJHMT,Cahill_et_al_JAP2003,Xuan_et_al_JAP2006,Desai_APL2011,Cahill_et_al_APR2013}. 

In the present work, we study the balance between the heat generated
by the NM subjected to the AMF and the heat lost to the sample's immediate
environment, a balance that turns out to determine the temperature
time profile observed in ferrofluids. For this purpose, we present
a phenomenological study of the heat exchange of a (non-interacting)
polydisperse assembly of NM, subjected to an external AMF, with its
environment. More precisely, we compute the heat $Q$ generated by
the nanomagnets (or the SAR $\mathfrak{S}$) through the dynamic response
given by the out-of-phase component of the AC susceptibility ($\chi^{\second}$)
of the system. Next, we describe the loss of this heat to the environment
with the help of a balance equation including a term that describes
a kind of Newton's law of cooling. The latter involves a phenomenological
coefficient $L_{n}$, the heat transfer coefficient, which we estimate
for two NM species, the maghemite and magnetite, using the experimental
data available for these materials and variable AMF characteristics
(amplitude and frequency) \citep{muraseEtal_rpt11,SHAH201596}. The
coefficient $L_{n}$ provides us with an additional means for assessing
the rate of heat losses to the environment and the corresponding results
obtained here, for various parameters of the NM assembly, could be
checked in future experiments. This could then help to develop a more
refined model for the mechanisms of heat generation and losses in
NM assemblies used in MH. 

We would like to add a final word regarding the comparison of our
model with experiments. First of all, we emphasize that our objective
here is not to build a model that would quantitatively fit the curves
measured for polydisperse assemblies of nanomagnets. This would not
be possible, anyway, given the various simplifications, exemplified
by the absence of inter-particle interactions. Instead, the main idea
to compare with experiments our results for the time profile of the
temperature elevation is twofold: 
\begin{enumerate}
\item Show that, at the time scales investigated in experiments, the observed
temperature profile is mostly determined by the balance equation that
accounts for i) the conversion into heat of the electromagnetic energy
brought in by the AMF and ii) its loss into the environment.
\item Estimate the order of magnitude of the Newton coefficient and investigate
its dependence on the AMF characteristics that are accessible to today's
measurements.
\end{enumerate}

\section{Model for heat generation and loss}

\begin{figure}[H]
\centering{}\centering{}\includegraphics[width=7cm]{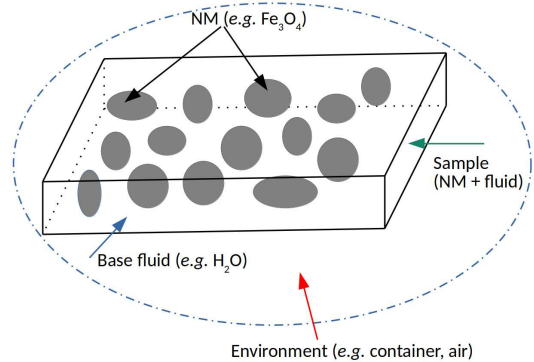}
\caption{\label{fig:NHCartoon}Sketch of the model system: a sample composed
of nanomagnets (NM) as heat sources, and the embedding fluid (base
fluid), coupled to its environment (air).}
\end{figure}

We start by defining the model system which we use for studying the
diffusion mechanisms of the heat generated by the nanomagnets (NM)
into their hosting medium and then to the environment. Our model system
is sketched in Fig. \ref{fig:NHCartoon}. It is composed of 
\begin{enumerate}
\item the sample (ferrofluid): an assembly of nanomagnets (magnetite, for
instance) floating in a fluid, the base fluid (bf), which is chosen
here to be water, 
\item the environment: may be the container of the sample (a tube, a box)
or air. 
\end{enumerate}
In the general scheme for SAR measurement, the sample is initially
set to the same temperature as its environment, $T_{0}$. Then, the
AMF is switched on at $t=0$. The latter excites the NM which then
release heat within the sample, \emph{i.e.} they heat up the fluid.
The heat thus generated then diffuses across the sample and the heat
flux reaches the sample's limits, or its interface with the environment,
to which the heat is transferred by conduction, convection or radiation.
If we assume that the NM are homogeneously distributed through the
sample, the heat power by unit volume is $\mathfrak{S}/V_{{\rm ff}}$,
where $V_{{\rm ff}}$ is the volume of the whole sample (NM + base
fluid), \emph{i.e.} the ferrofluid.

\subsection{Heat balance equation }

The time evolution of the temperature within the sample, measured
at position $\bm{r}$, is governed by the heat-diffusion equation
with sources, namely \citep{AndreuNatividad_ijh13} 
\begin{equation}
\rho\:c_{v}\frac{\partial T}{\partial t}\left(\bm{r},t\right)-\kappa\bm{\nabla}^{2}T\left(\bm{r},t\right)=\mathfrak{\mathcal{P}}/V_{s},\label{eq:HeatEqt}
\end{equation}
where $\rho\left({\rm kg}/{\rm m}^{3}\right)$, $c_{v}\left({\rm J/K/kg}\right)$,
and $\kappa\left({\rm W/K/m}\right)$ are respectively the density,
the specific heat and the thermal conductivity of the sample, all
assumed to be uniform. $\mathfrak{\mathcal{P}\left({\rm W}\right)}$
is the heat power and $V_{s}$ the sample volume. In this case, Eq.
(\ref{eq:HeatEqt}) can be solved analytically, but in practice it
is more efficient to make use of optimized numerical algorithms. For
the calculation of the SAR, a simpler model may be derived if one
ignores the temperature gradients within the sample, upon which the
spatial derivative of $T$ is dropped from the equation above. Obviously,
this temperature gradient is what conveys the heat generated by the
NM to the environment. Nevertheless, this assumption may apply if
the (thermal) relaxation time within the sample is much shorter than
the relaxation time of heat diffusion through the sample-environment
interface. In practice, this may be achieved by using a highly conductive
sample with a weak link to its environment. Within this approximation,
one may derive the temporal evolution of the temperature within the
sample from the power balance between the sample (NM + BF) and its
environment. Accordingly, upon dropping the gradient term from Eq.
(\ref{eq:HeatEqt}) and adding the contribution of the heat flow (power
loss) from the sample into the environment, we obtain the following
balance equation (upon multiplying by the sample volume $V_{{\rm ff}}$)
\begin{equation}
C_{{\rm ff}}\frac{dT}{dt}=\mathfrak{\mathcal{P}}\left(T\right)-L_{{\rm ff}}\left[T\left(t\right)-T_{0}\right].\label{eq:TBalance}
\end{equation}

Here $C_{{\rm ff}}=\sum_{i}c_{i}m_{i}$ is the heat capacity (in J/K)
of the sample. We have introduced the coefficient $L_{{\rm ff}}$(W/K)
as the analog of the heat transfer coefficient between the sample
and its environment. This equation implies that the temperature rise
in the ferrofluid is the result of a balance between the heat power
generated by the NM (within the ferrofluid) and the heat power lost
to the environment. The last process is described here by the so-called
Newton's law of cooling {[}see the textbook \citep{BirdEtAl_JWS07}
for a more general discussion{]}. At the initial time $t=0$, when
the AC field is switched on, we have $T\left(t=0\right)=T_{0}$, which
is the temperature of the sample. Note that $\mathfrak{\mathcal{P}}\left(T\right)$
in Eq. (\ref{eq:TBalance}) is the total heat power generated by the
entire assembly of NM. Hence, if one computes the SAR for a single
NM, one has to multiply by the total number of NM in the ferrofluid.

In Ref. \citep{AndreuNatividad_ijh13}, $C_{{\rm ff}},\mathfrak{\mathcal{P}}$
and $L_{{\rm ff}}$ were assumed to be constant upon which the equation
above is analytically solved as follows 
\begin{equation}
T\left(t\right)=T_{0}+\Delta T_{{\rm max}}\left(1-e^{-t/\varsigma}\right)\label{eq:TSol-constantCoeffs}
\end{equation}
where $\Delta T_{{\rm max}}=\mathfrak{\mathcal{P}}/L_{{\rm ff}}$
and $\varsigma=C_{{\rm ff}}/L_{{\rm ff}}$. Thus, in the steady state
($t\rightarrow\infty$), the heat generated and that lost become equal
and the sample temperature remains constant at $T_{{\rm max}}=T_{0}+\Delta T_{{\rm max}}$.

\subsection{SAR from linear response}

Next, we discuss the calculation of the SAR. This may be inferred
either from the area of the dynamic hysteresis loop or from the dynamic
response of the system represented by the AC susceptibility. A detailed
discussion can be found in Refs. \citep{dejardin_etal_SAR_JAP_2017,DejVerKach_springer20,DejVerKach_jap20}
and references therein. In both cases, one is faced with a nonlinear
problem which makes it rather difficult, if not impossible, to come
up with an analytical solution. However, the general situation can
be tackled with the help of a numerical algorithm that could proceed
as follows: the Landau-Lifshitz-Langevin equation can be solved \citep{garlaz98prb}
at a given temperature $T$ to obtain the AC susceptibility and thereby
the SAR. Next, the latter is used in (\ref{eq:TBalance}) to solve
it for $T\left(t\right)$, which in turn is plugged into Landau-Lifshitz-Langevin
equation and so on. The whole process should converge to the solution
$\mathfrak{S}$ and $T$. Such a numerical procedure will be followed
in a future work. In the situation of interest to us here, the SAR
$\mathfrak{S}$ (or power $\mathfrak{\mathcal{P}}$), is obtained
from an analytical expression of the AC susceptibility $\chi_{{\rm ac}}$,
see \citep{dejardin_etal_SAR_JAP_2017,DejVerKach_springer20,DejVerKach_jap20}.
However, the temperature enters in a rather involved way, explicitly
through the coefficients of $\chi_{{\rm ac}}$ and, implicitly, through
the relaxation time nested in $\chi_{{\rm ac}}$. Therefore, if we
consider the nonlinear corrections to $\chi_{{\rm ac}}$, as was done
in Ref. \citep{DejVerKach_jap20}, the differential equation (\ref{eq:TSol-constantCoeffs})
can only be numerically solved. In fact, in the general situation,
even if only the linear contribution is used in $\chi_{{\rm ac}}$,
no simple analytical expression can \emph{a priori} be derived for
the time rise of temperature $T\left(t\right)$. In this case, we
may content ourselves with a numerical solution of Eq. (\ref{eq:TBalance})
upon inserting an analytical expression for $\chi_{{\rm ac}}$ given
in Refs. \citep{dejardin_etal_SAR_JAP_2017,DejVerKach_springer20,DejVerKach_jap20},
which we summarize now.

The SAR $\mathfrak{S}$ of a nanomagnet, of volume $V$ and magnetic
moment $m=M_{s}V$, subjected to an AC magnetic field $H\left(t\right)=h\cos\left(\omega t\right)$
is given by \citep{rosensweig02j3m,dejardin_etal_SAR_JAP_2017} 
\begin{equation}
\mathfrak{S}\left(T\right)=\frac{\mu_{0}h^{2}}{2}\omega\chi^{\prime\prime}\left(T\right)\label{eq:SAR-Chi2}
\end{equation}
where the out-of-phase component of the AC susceptibility $\chi=\chi^{\prime}-i\chi^{\prime\prime}$
reads (in the Debye approximation) 
\begin{equation}
\chi^{\prime\prime}\left(T\right)=\chi_{{\rm eq}}\frac{\eta}{1+\eta^{2}}\label{eq:ChiIm}
\end{equation}
with the equilibrium susceptibility 
\begin{equation}
\chi_{{\rm eq}}\left(T\right)=\frac{\mu_{0}m^{2}}{3k_{B}T}.\label{ChiEqui}
\end{equation}

$\eta$ is the reduced relaxation time, \emph{i.e.}, $\eta\equiv\omega\tau_{{\rm eff}}$,
where $\tau_{{\rm eff}}$ is the effective relaxation time given by
\begin{equation}
\tau_{{\rm eff}}^{-1}=\tau_{{\rm N}}^{-1}+\tau_{{\rm B}}^{-1}\label{eq:TauEff}
\end{equation}
with $\tau_{{\rm N}}$ being the Néel relaxation time related with
the motion of the net magnetic moment in the energy potential of the
NM and $\tau_{{\rm B}}$, the Brown relaxation time related with the
physical motion of the NM within the fluid. More precisely, in zero
DC magnetic field and using the usual notation, we have 
\begin{equation}
\omega\tau_{{\rm N}}=\frac{\sqrt{\pi}}{2}\frac{\omega\tau_{0}}{\sqrt{\sigma}}\,e^{\sigma},\quad\sigma\left(T\right)=\frac{KV}{k_{B}T}\label{eq:RelaxTime}
\end{equation}
according to the Arrhenius law and \citep{raishl94acp} 
\[
\tau_{{\rm B}}=\frac{3\xi V_{H}}{k_{B}T}
\]
where $\xi$ is the viscosity of the medium (fluid); for maghemite,
$\xi=0.00235\,{\rm kg/m/s}$ (or $0.00235\,{\rm Pa.s}$). $V_{H}$
is the hydrodynamic volume that is larger than the magnetic volume
$V=4\pi R^{3}/3$ of a NM of radius $R$. In general, $V_{H}$ is
much larger than $V$ \citep{muraseEtal_rpt11,OGRADY_JPHYSD2013},
since the hydrodynamic radius may be up to $50\,{\rm nm}$ more than
the magnetic one.

Summarizing, we have

\begin{equation}
\mathfrak{S}\left(T\right)=\frac{\mu_{0}h^{2}}{2}\omega\chi_{{\rm eq}}\left(T\right)\frac{\eta\left(T\right)}{1+\eta\left(T\right)^{2}}.\label{eq:SAR-in-T}
\end{equation}

In order to give estimates of the various physical parameters, we
have to give each term appearing in Eq. (\ref{eq:TBalance}) a clear
meaning. Accordingly, for a given magnetic substance, most often maghemite
or magnetite, the power $\mathfrak{\mathcal{P}}$ divided by the mass
of the magnetic substance ($m_{{\rm Fe}}$) yields the SAR $\mathfrak{S}$
in W/g. Then, the constant $C_{{\rm ff}}$ in Eq. (\ref{eq:TBalance}),
which is defined by $C_{{\rm ff}}=m_{{\rm bf}}c_{v}\left({\rm bf}\right)+m_{{\rm Fe}}c_{v}\left({\rm Fe}\right)$,
will be replaced by the specific heat $c_{{\rm ff}}\equiv C_{{\rm ff}}/m_{{\rm Fe}}$which
is measured in ${\rm J/K/g}$ while $\left[C_{{\rm ff}}\right]={\rm J/K}$.
Likewise, we define the coefficient $l_{{\rm ff}}\equiv L_{{\rm ff}}/m_{{\rm Fe}}$
($\left[l_{{\rm ff}}\right]={\rm W/K/g}$). Note that the specific
heat $c_{{\rm ff}}$ may also be rewritten as \emph{ 
\begin{align}
c_{{\rm ff}} & =c_{v}\left({\rm Fe}\right)+\frac{\rho_{{\rm bf}}}{\rho_{{\rm Fe}}}c_{v}\left({\rm bf}\right).\label{eq:SpecificHeat}
\end{align}
}where $\rho_{{\rm bf}}$ is the density of the base fluid and $\rho_{{\rm Fe}}$
the concentration of the magnetic substance in the fluid.

\subsection{Balance equation and temperature profile}

Now, comes the discussion of the coefficient $l_{{\rm ff}}$, which
is measured in ${\rm W/K/g}$. This should be interpreted as the analog
of Newton's coefficient of heat transfer between the whole ferrofluid
and its environment which may be the container or air. This is an
unknown parameter that depends on the sample (its shape, size, etc),
and for which it is rather difficult to come up with a simple and
precise expression. So, in the present context, it is regarded as
a free parameter that could be estimated from experiment. Let us,
however, attempt to estimate its value and limits. At initial time,
$t=0$, or say at short times, the heat just generated by the nanomagnets
diffuses through the ferrofluid, without yet reaching out to the environment.
At this stage, we may set $l_{{\rm ff}}=0$ and obtain the equation
\begin{equation}
c_{{\rm ff}}\left.\frac{dT}{dt}\right|_{t\simeq0}=\mathfrak{S}\left(T\right).\label{eq:ShortTimes}
\end{equation}

This is the equation used in experiments \citep{SHAH201596,AndreuNatividad_ijh13}
to infer the SAR from the measurements of the initial slope of temperature
elevation. As time evolves, the heat generated and diffused within
the ferrofluid starts to ``trickle'' into the environment. This
means that the coefficient $L_{{\rm ff}}$ (or $l_{{\rm ff}}$) acquires
nonzero values, and eventually the system (ferrofluid + its environment)
reaches a steady state in which the temperature of the ferrofluid
is governed by Eq. (\ref{eq:TSol-constantCoeffs}), assuming that
$L_{{\rm ff}}$ does not depend on $T$. As $t\rightarrow\infty$,
the amount of heat generated within the ferrofluid and that lost to
the environment become equal and the sample temperature remains constant
at $T_{{\rm max}}=T_{0}+\Delta T_{{\rm max}}$ with $\Delta T_{{\rm max}}=\mathfrak{S}/l_{{\rm ff}}$.
Hence, the heat-transfer coefficient $l_{{\rm ff}}$ my be inferred
from the asymptotic value of the temperature rise and the corresponding
SAR observed in experiments.

Now, with the newly defined coefficients, Eq. (\ref{eq:TBalance})
becomes 
\begin{equation}
c_{{\rm ff}}\frac{dT}{dt}=\mathfrak{S}-l_{{\rm ff}}\left[T\left(t\right)-T_{0}\right].\label{eq:TBalance-sar}
\end{equation}

However, it is more convenient to work with a dimensionless equation.
For this purpose, we introduce the time scaling factor (similar to
a ``relaxation time''), $\varsigma=c_{{\rm ff}}/l_{{\rm ff}}$ and
thereby the dimensionless time $\tau\equiv t/\varsigma$. Next, we
define the relative temperature rise $\theta\left(t\right)\equiv\left(T-T_{0}\right)/T_{0}$
and the dimensionless SAR $\Xi\equiv\mathfrak{S}/\left(T_{0}l_{{\rm ff}}\right)$
so that Eq. (\ref{eq:TBalance-sar}) yields the final form of the
equation we have to solve in order to obtain the temperature time
profile

\begin{equation}
\frac{d\theta}{d\tau}=-\theta\left(\tau\right)+\Xi\left[\theta\left(\tau\right)\right].\label{eq:DeltaT-dimless}
\end{equation}

Note that for the numerical solution of this equation, the variable
$T$ in the expression of SAR given in Eq. (\ref{eq:SAR-in-T}) has
to be replaced by $T=T_{0}\left(1+\theta\right)$.

In our discussion above regarding the solution of Eq. (\ref{eq:TBalance}),
or (\ref{eq:DeltaT-dimless}), we emphasized the fact that, in the
most general situation, it is not easy to obtain an analytical expression
for the time profile of the temperature elevation. However, in practice
we note that for maghemite, for instance, $\theta_{{\rm max}}=\left(\Delta T\right)_{{\rm max}}/T_{0}$
is about $25/318\simeq0.08$ and as such we may analytically solve
Eq. (\ref{eq:DeltaT-dimless}) upon expanding $\Xi\left(\theta\right)$
to first order in $\theta$. In order to illustrate this in a simple
manner, we ignore the contribution of the Brownian motion to the relaxation
time in Eq. (\ref{eq:TauEff}). Doing so, we obtain the solution 
\begin{equation}
\theta\left(\tau\right)=\frac{\Xi\left(T_{0},h,\omega\right)}{\Gamma\left(T_{0},h,\omega\right)}\left[1-e^{-\Gamma\left(T_{0},h,\omega\right)\tau}\right]\label{eq:TProfile-Analytic}
\end{equation}
with $\Xi\left(T_{0},h,\omega\right)$ being simply the dimensionless
SAR evaluated at $T_{0}$ and 
\[
\Gamma\left(T_{0},h,\omega\right)=1+\Xi\left(T_{0},h,\omega\right)\Omega\left(T_{0},\omega\right)
\]
where 
\[
\Omega\left(T_{0},\omega\right)=\frac{\left(1+2\sigma_{0}\right)+\left(3-2\sigma_{0}\right)\eta_{0}^{2}}{2\left(1+\eta_{0}^{2}\right)},\quad\sigma_{0}=\frac{KV}{k_{B}T_{0}}
\]
and $\eta_{0}\equiv\eta\left(T_{0}\right)$, see Eq. (\ref{eq:RelaxTime}).
The temperature elevation $\Delta T$ is analytically obtained from
Eq. (\ref{eq:TProfile-Analytic}) upon multiplying by $T_{0}$. It
is straightforward to extend this solution using the effective relaxation
time $\tau_{{\rm eff}}$ instead of $\tau_{{\rm N}}$.

Note that in practice, for the physical parameters of prototypical
ferrofluids used in MH, the second term in $\Gamma\left(T_{0},h,\omega\right)$
turns out to be a very small number. As such, $\theta\left(\tau\right)\propto\left(1-e^{-\tau}\right)$,
which goes to $1$ for long times leading to the saturation of $\Delta T$
observed in experiments \citep{muraseEtal_rpt11,SHAH201596} and discussed
above. This is indeed the profile that obtains from Eq. (\ref{eq:TSol-constantCoeffs})
when rewritten in the same notation as Eq. (\ref{eq:TProfile-Analytic}),
namely $\theta\left(t\right)=\left(T-T_{0}\right)/T_{0},\Delta T_{{\rm max}}/T_{0}=\theta_{{\rm max}}$,
and $\tau=t/\varsigma$, leading to $\theta\left(t\right)=\theta_{{\rm max}}\left(1-e^{-\tau}\right)$.
This comparison establishes the connection between our approach and
the ``asymptotic'' time behavior in Eq. (\ref{eq:TSol-constantCoeffs}).
In particular, we see that, in a more general situation, both the
rate of temperature elevation, $\Gamma\left(T_{0},h,\omega\right)$,
and the prefactor, $\Xi\left(T_{0},h,\omega\right)/\Gamma\left(T_{0},h,\omega\right)$,
depend in a non trivial way on the AMF amplitude and frequency and
the sample temperature.

\begin{figure}[H]
\begin{centering}
\includegraphics[scale=0.22]{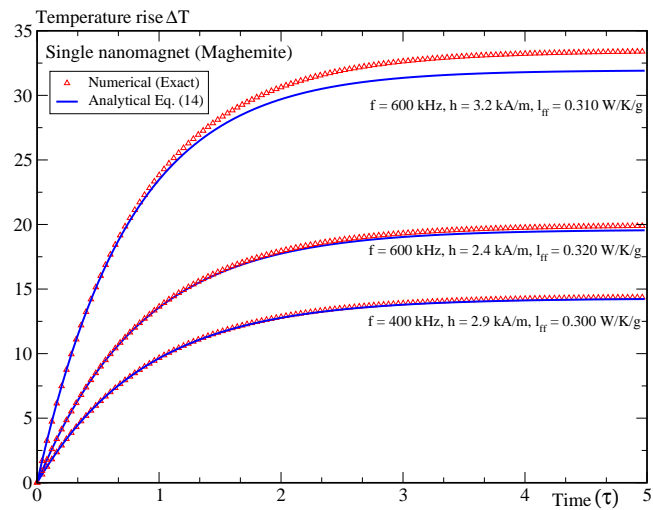}
\par\end{centering}
\caption{\label{fig:DT-AnaNum_vfvh}Time profile of the temperature elevation
in a single NM: a comparison of the numerical solution of Eq. (\ref{eq:DeltaT-dimless}),
in symbols, with the analytical approximation (\ref{eq:TProfile-Analytic}),
in full blue lines. The physical parameters are taken from Fig. \ref{fig:MuraseFit}
for maghemite assuming a monodisperse assembly.}
\end{figure}

In Fig. \ref{fig:DT-AnaNum_vfvh}, we compare the results from the
numerical (exact) solution of Eq. (13) to the analytical approximation
in Eq. (14), using the data for maghemite as in Fig. \ref{fig:MuraseFit}
and assuming a monodisperse assembly of nanomagnets with a diameter
of $15.2$ nm (see next Section). We see that the analytical expression
in Eq. (\ref{eq:TProfile-Analytic}) is a rather good approximation
for relatively low frequency and/or small amplitude of the AMF. However,
the experimental data in Figs. \ref{fig:ShahFit} and \ref{fig:MuraseFit}
(for magnetite and maghemite) are only available up to $\tau\sim0.6$,
and in this range the analytical and the numerical solutions cannot
be distinghuised on the scale used for the plots in Fig \ref{fig:DT-AnaNum_vfvh}.
We could have included the curves from the asymptotic behaviour $\theta\left(t\right)=\theta_{{\rm max}}\left(1-e^{-\tau}\right)$,
from Eq. (\ref{eq:TSol-constantCoeffs}), but for this we would need
to manually insert, for each set of physical parameters, the asymptotic
value $\theta_{{\rm max}}$ if it is known. 

We may conclude then that the analytical profile in Eq. (\ref{eq:TProfile-Analytic})
recovers very well the time profile of the temperature elevation in
single nanomagnets, as obtained from the balance equation (\ref{eq:DeltaT-dimless}).
This comparison is given here for the example of maghemite but should
apply to any substance as long as the condition that $\Delta T/T_{0}$
remains small for the linear approximation underlying Eq. (\ref{eq:TProfile-Analytic})
to be valid. Obviously, the same agreement should obtain for a dilute
assembly as well. 

Incidentally, we already see that the increase of the AMF amplitude
has a stronger effect than that of increasing its frequency {[}see
further discussion in the next Section{]}.

\section{\label{sec:Results-and-discussion}Effects of field amplitude and
frequency: comparison with experiments}

We have (numerically) solved the differential equation (\ref{eq:DeltaT-dimless})
and derived the time profile of the temperature rise within the ferrofluid
for both a single NM and an assembly thereof, using the effective
relaxation time $\tau_{{\rm eff}}$ given by Eq. (\ref{eq:TauEff}).
We have calculated the temperature profile $T\left(t\right)$ upon
varying the relevant physical parameters; we have considered the two
examples of maghemite and magnetite for which experimental data are
available. For the assembly, we consider the case closest to the samples
investigated for MH, namely polydisperse assemblies with randomly
distributed anisotropy axes; the volume distribution is given by the
log-normal law 
\[
W\left(V,\delta\right)=\frac{1}{\delta\sqrt{2\pi}V}e^{-\log^{2}\left(V/V_{{\rm m}}\right)/\left(2\delta^{2}\right)}.
\]
with the mean volume $V_{{\rm m}}$ and standard deviation $\delta$.
Then, the SAR of the assembly is given by the volume average ($W$
is already normalized) 
\begin{equation}
\overline{\mathfrak{S}}\left(\omega,T,h\mid V_{{\rm m}},\delta\right)=\int W\left(V,\delta\right)\mathfrak{S}\left(\omega,T,h,V\right)\,dV.\label{eq:SARAve}
\end{equation}

In the following sections, we will discuss our results and compare
them with experiments performed on two prototypical NM assemblies
most often used in MH applications, namely assemblies of maghemite
and magnetite NM studied in Refs. \citep{muraseEtal_rpt11} and \citep{SHAH201596},
respectively. For later use, we now summarize the physical parameters
for these systems.

For magnetite, we will mainly refer to and compare with the data provided
by Shah et al. \citep{SHAH201596}, namely nanoparticles with iron
oxide concentration of $8.6$mg/cm$^{3}$ and a diameter of $9$nm;
an anisotropy constant $K=3\times10^{4}\mathrm{J}/\mathrm{m^{3}}$,
a saturation magnetization $M_{s}=480\times10^{3}\mathrm{A/m}$, and
a density $\rho=5.2\mathrm{g/\mathrm{cm^{3}}}$. The anisotropy field
is taken equal to $0.96K/M_{s}$, due to the random distribution of
easy axes.

Let us now discuss the specific heat and ``relaxation time'' defined
respectively in Eqs. (\ref{eq:SpecificHeat}) and (\ref{eq:DeltaT-dimless}).
Using the data for magnetite in water, as studied by Shah \emph{et
al.} \citep{SHAH201596}, we have: $c_{v}\left({\rm bf}\right)=c_{v}\left({\rm water}\right)\simeq4.184\,{\rm J/K/g}$
and $c_{v}\left({\rm Fe}\right)=c_{v}\left({\rm magnetite}\right)\simeq3.8\times10^{3}\,{\rm J/K/L}$,
or using the density of magnetite ($5.2\,{\rm g/cm^{3}}$), we get
$c_{v}\left({\rm Fe}\right)\simeq0.731\,{\rm J/K/g}$. These parameters
of the two ferrofluid components are independent of the sample studied
in MH experiments. Now, in a given experiment one uses a specific
mixture (or ferrofluid), \emph{i.e.} with a given ratio $m_{{\rm bf}}/m_{{\rm Fe}}$
or $\rho_{{\rm bf}}/\rho_{{\rm Fe}}$. Shah \emph{et al.} \citep{SHAH201596}
used $1.2\,{\rm mL}$ of magnetite ferrofluid (magnetite in water)
and a concentration $\rho_{{\rm Fe}}\simeq8.6\,{\rm mg/mL}\simeq8.6\,{\rm kg/m}^{3}$
of magnetite in the ferrofluid. From this we infer the ferrofluid
specific heat $c_{v}\left({\rm ferrofluid}\right)\equiv\left[m_{{\rm bf}}c_{v}\left({\rm bf}\right)+m_{{\rm Fe}}c_{v}\left({\rm Fe}\right)\right]/m_{{\rm total}}\simeq4.156\,{\rm J/K/g}$,
which is slightly smaller than that of water. This is reasonable knowing
that the amount of magnetic substance is very small compared to the
total mass of the ferrofluid. The mass of Fe is given by $m_{{\rm Fe}}=\rho_{{\rm Fe}}V_{{\rm ff}}\simeq1.032\times10^{-2}\,{\rm g}$.
Finally, the constant $c_{{\rm ff}}$ for the sample studied by Shah
\emph{et al}. \citep{SHAH201596} evaluates to $c_{{\rm ff}}\simeq486.43\,{\rm J/K/g}$.

For maghemite, we use the data for the samples studied by Murase et
al. \citep{muraseEtal_rpt11}. These are superparamagnetic iron oxide
particles (Resovist) coated with carboxydextran with a volume fraction
of $\phi=0.0035$ (or a concentration of 5.58 mg/cm$^{3}$) and a
diameter of $15.2$ nm. The magnetic characteristics of maghemite
are : $K=0.47\times10^{4}\mathrm{J}/\mathrm{m^{3}},$ $M_{s}=414\times10^{3}\mathrm{A/m,}$
$\rho=4.9$ g/cm$^{3}$. Then, we obtain for the maghemite ferrofluid
$c_{{\rm ff}}\simeq750.4\,{\rm J/K/g}$.

\subsection{Effect of AMF amplitude and frequency}

\begin{figure*}
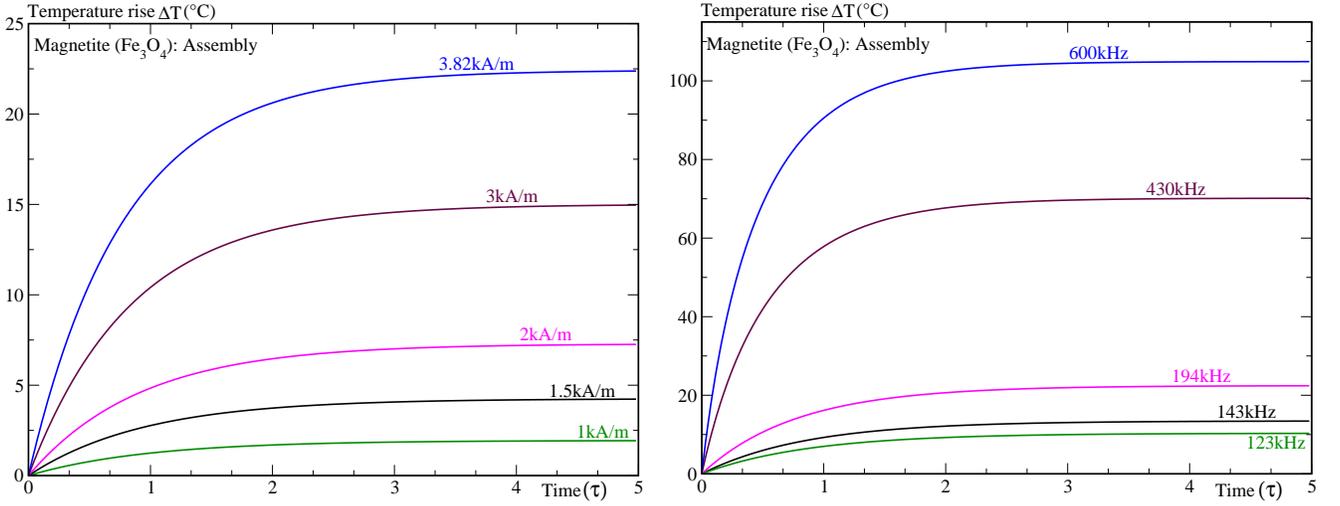

\begin{centering}
\includegraphics[scale=0.22]{DT-Magnetite-Ass_vH}\quad{}\includegraphics[scale=0.22]{DT-Magnetite-Ass} 
\par\end{centering}
\caption{\label{fig:deltaT(t)Magnetite}Time profile of temperature rise in
magnetite NM with $D_{{\rm m}}=9\,{\rm nm},\delta=0.25$. (Left) varying
the AMF amplitude at fixed frequency ($f=194\,{\rm kHz}$) and (right)
varying the AMF frequency at fixed amplitude ($h=38.2\,{\rm kA/m}$).}
\end{figure*}

In Fig.\ref{fig:deltaT(t)Magnetite}, we plot the temperature rise
as obtained from Eq. (\ref{eq:DeltaT-dimless}) using the data for
magnetite ferrofluids. On the left, the results are for a varying
frequency of the AMF with fixed amplitude ($h=38.2$kA/m) and, on
the right, for a varying amplitude and fixed frequency \textcolor{black}{($f=194$
kHz).} Note that the highest values of the temperature elevation correspond
to temperatures that exceed the water boiling point. In fact, we have
extended the time range solely to see that the temperature profile
formally reaches an asymptote.

These temperature profiles agree very well with those observed in
experiments (see below for a quantitative comparison). We have seen
that, for both maghemite and magnetite, the asymptotic values of the
temperature elevation $\Delta T$ are higher for the assembly of nanoparticles
than for a single particle (not shown here), but this obviously depends
on $D_{{\rm m}}$ and $\delta$.

For the two substances, we see that the initial slope of the temperature
rise, \emph{i.e.} $\left.\Delta T/\Delta t\right|_{t=0}$, increases
with the AMF frequency as was observed in experiments \citep{muraseEtal_rpt11}.
This stems from the fact that $\mathfrak{S}\sim\omega^{2}$ (for high
frequencies), see Eq. (\ref{eq:SAR-in-T}) remembering that $\eta\sim\omega$.
Similarly to the frequency effect, the maximum temperature elevation
increases with the AMF amplitude ($\mathfrak{S}\sim h^{2}$). The
effect of the AMF amplitude is, however, stronger than that of its
frequency because, in fact, the latter also appears in the denominator
of $\mathfrak{S}$ since $\mathfrak{S}\sim\omega^{2}/\left(1+\tau_{{\rm eff}}^{2}\omega^{2}\right)$,
see Eq. (\ref{eq:SAR-in-T}).

\subsection{Heat transfer coefficient}

\begin{figure*}[t]
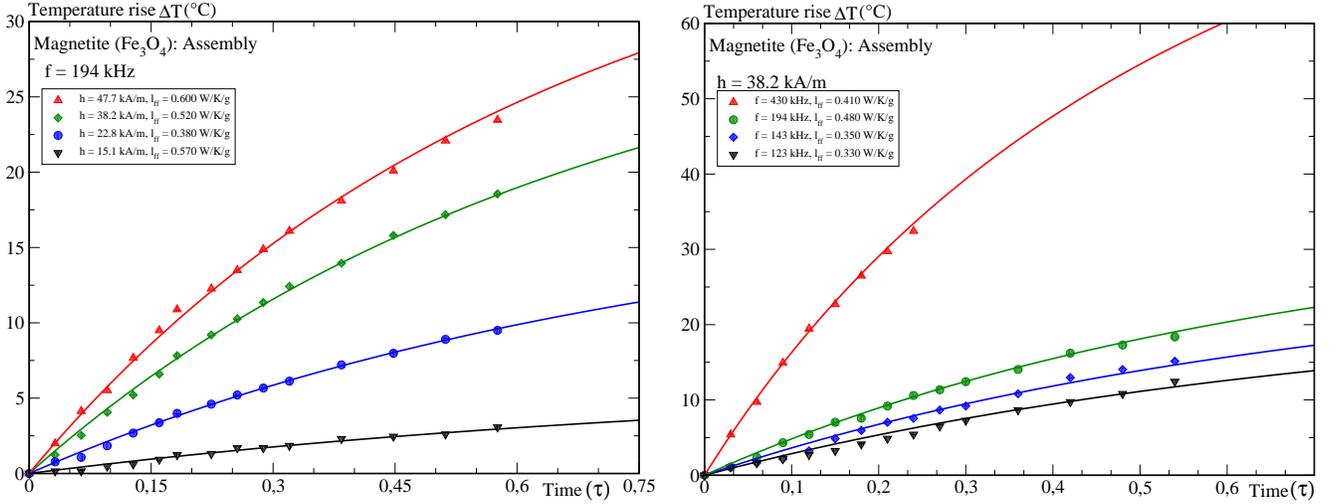

\begin{centering}
\includegraphics[scale=0.22]{DT-Magnetite-Ass_ShahFit}\quad{}\includegraphics[scale=0.22]{DT-Magnetite-Ass_ShahFit_freq} 
\par\end{centering}
\caption{\label{fig:ShahFit} Time profile of the temperature rise i) in symbols,
experimental data for magnetite (Figs. 4 \& 5, Ref. \citep{SHAH201596})
and ii) in full lines, as obtained from Eq. (\ref{eq:DeltaT-dimless}).
(Left) for different values of the AMF amplitude and $f=194$kHz and
(right) for different values of its frequency and $h=38.2$kA/m. The
heat transfer coefficient $l_{{\rm ff}}$ is indicated for each (best)
fit.}
\end{figure*}

Now, we discuss the coefficient of heat transfer $l_{{\rm ff}}$.

In the general Newton's law of cooling at a solid-fluid interface
\citep{BirdEtAl_JWS07}, the heat flux may be related to the difference
between the solid surface temperature $T_{s}$ and the bulk fluid
temperature $T_{0}$: $Q=h\left(T_{s}-T_{0}\right)$ and the coefficient
$h$ depends on various parameters. For example, in the case of a
fluid flowing around submerged objects, \emph{e.g. }spheres of radius
$R$, the coefficient $h$ is proportional to $R^{2}$. The study
of the specific case of convective heat transfer in magnetic nanofluids,
\emph{e.g.} Fe$_{2}$O$_{3}$/water, is a very active research which
aims at a more precise characterization of the heat transfer coefficient
as a function of the applied magnetic field and the NM assembly properties
\citep{AmrollahiEtAl_ichmt10,ShaEtAl_ate17,EsmaeiliEtAl_ate17}. For
example, it is shown that the (convective) heat transfer coefficient
is enhanced in the presence of NM and is strongly dependent on the
magnetic field strength.

\begin{figure*}[t]
\begin{centering}
\includegraphics[scale=0.22]{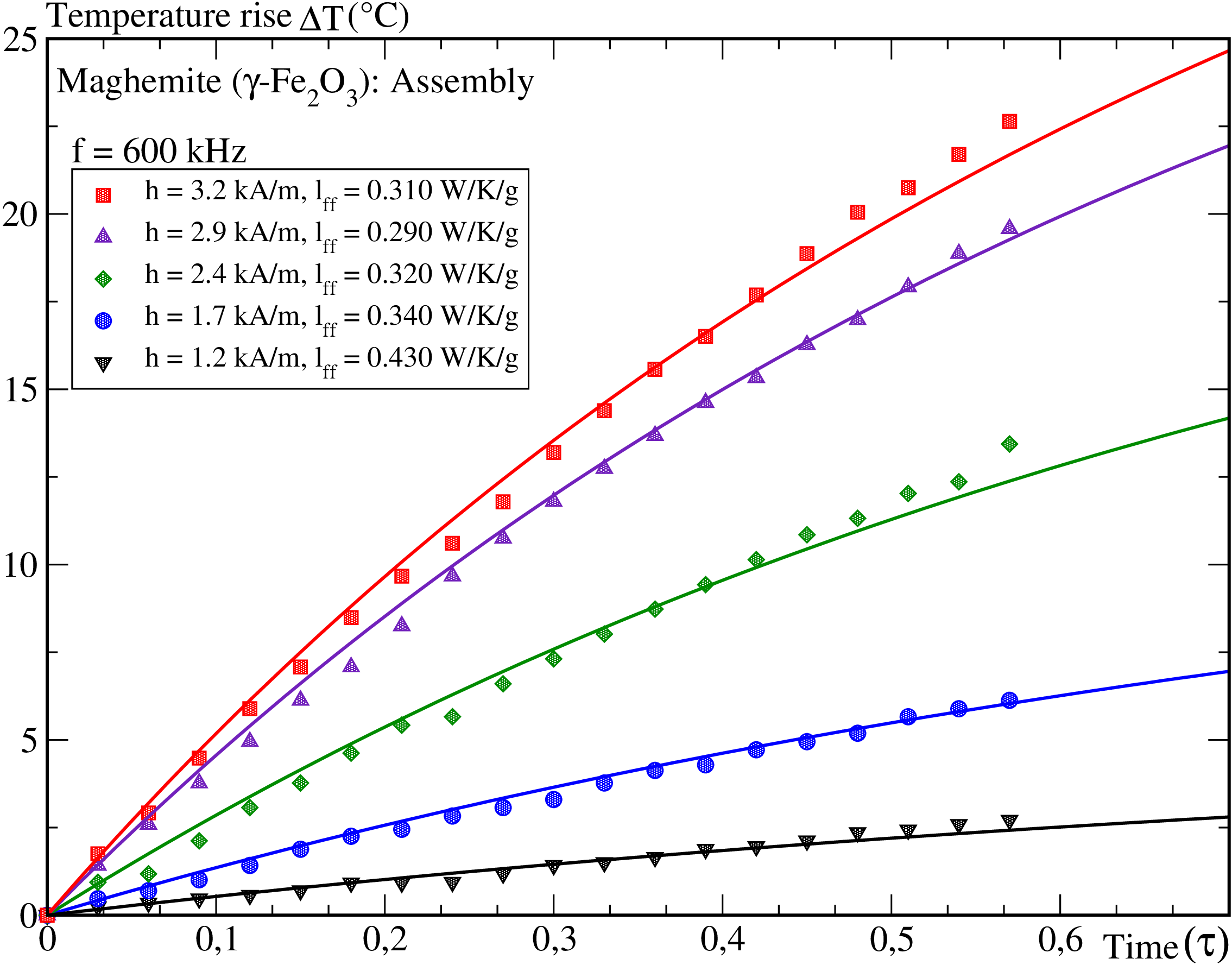}\quad{}\includegraphics[scale=0.22]{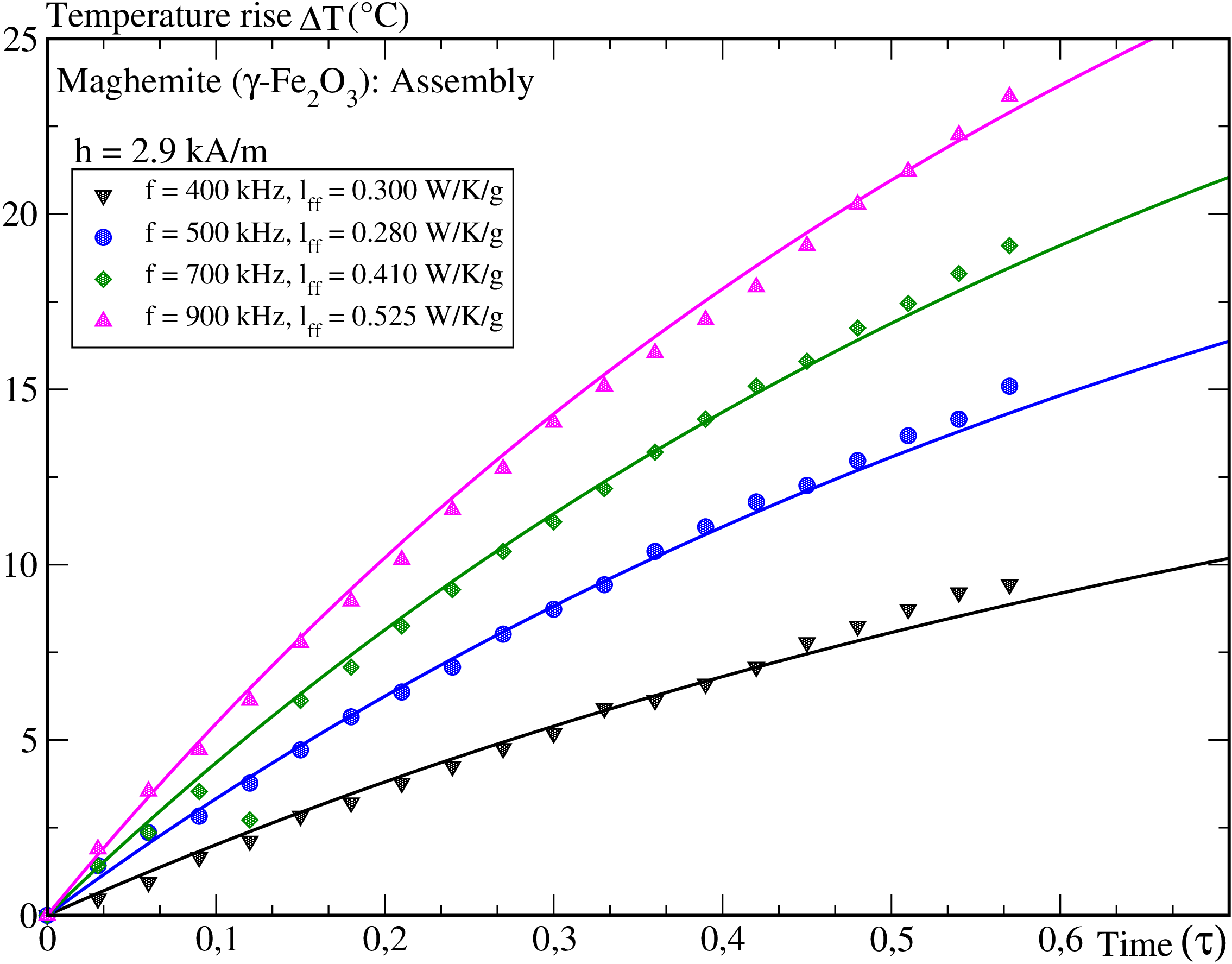} 
\par\end{centering}
\caption{\label{fig:MuraseFit} Time profile of the temperature rise i) in
symbols, experimental data for maghemite (Figs. 7a \& 8a, Ref. \citep{muraseEtal_rpt11})
and ii) in full lines, as obtained from Eq. (\ref{eq:DeltaT-dimless}).
(Left) for different values of the AMF amplitude and $f=600$kHz and
(right) for different values of its frequency and $h=2.9$kA/m. The
heat transfer coefficient $l_{{\rm ff}}$ is indicated for each (best)
fit.}
\end{figure*}

In analogy with Newton's law of cooling, we have introduced the heat-transfer
coefficient $l_{{\rm ff}}$ in the balance equation (\ref{eq:TBalance})
in order to model the heat transfer at the interface between the ferrofluid
and its environment. In the present study, $l_{{\rm ff}}$ is a phenomenological
parameter and the approach used here to build the balance equation
(\ref{eq:TBalance}) does not allow for a derivation of the behavior
of $l_{{\rm ff}}$ as a function of the system and excitation characteristics.
Nevertheless, a comparison of the results rendered by this approach
with the experimental data available to us today should help us infer
useful hints for a more elaborate investigation of this coefficient
and its characterization in MH ferrofluids.

Accordingly, in Figs \ref{fig:ShahFit} and \ref{fig:MuraseFit} we
plot (in full curves) the time profiles of the temperature rise obtained
from Eq. (\ref{eq:DeltaT-dimless}), for varying AMF amplitude and
frequency. The symbols are the experimental data reported in Refs.
\citep{SHAH201596} for magnetite and in \citep{muraseEtal_rpt11}
for maghemite ferrofluids. Overall, we see that the time profile of
temperature rise in ferrofluids, such as $\gamma$-Fe$_{2}$O$_{3}$
and Fe$_{3}$O$_{4}$ in water, is well described by the balance equation
(\ref{eq:DeltaT-dimless}). This good quantitative agreement is made
possible by an adjustment of the only ``free'' parameter provided
by the (phenomenological) heat-transfer coefficient $l_{{\rm ff}}$.
This agreement is also supported by the value of the simplest coefficient
of determination defined by $R^{2}=1-\Sigma_{\text{res}}/\Sigma_{\mathrm{tot}}$,
where $\Sigma_{\text{res}}=\sum_{i=1}^{\mathcal{N}}\left[\left(\Delta T\right)_{\mathrm{exp}}^{i}-\left(\Delta T\right)_{\mathrm{theo}}^{i}\right]^{2}$
is the residual sum of squares and $\Sigma_{\mathrm{tot}}=\sum_{i=1}^{\mathcal{N}}\left[\left(\Delta T\right)_{\mathrm{exp}}^{i}-\overline{\Delta T}\right]^{2}$
the total sum of squares, with $\overline{\Delta T}=\left(1/\mathcal{N}\right)\sum_{i=1}^{\mathcal{N}}\left(\Delta T\right)_{\mathrm{exp}}^{i}$.
Indeed, for the curves in Fig. \ref{fig:ShahFit} (magnetite), for
instance, we obtain $R^{2}\sim0.99$ for almost all curves ; we have
also estimated that an uncertainty of $1K$ in the temperature elevation
corresponds to $\Delta l_{{\rm ff}}\simeq10^{-2}$.

A more stringent goodness-of-fit test could be performed if more experimental
data were available for the AMF amplitude and frequency and the model
extended to interacting assemblies. Nevertheless, the adjustment we
perform here allows us to obtain an estimate of the coefficient $l_{{\rm ff}}$
for magnetite and maghemite ferrofluids and for different values of
AMF amplitude and frequency. In a more elaborate theory of heat transfer
in ferrofluids and their interface with the external environment,
the Newton heat-transfer coefficient should depend on the features
of the system (NM+bf) as well as the external stimulus. However, in
the present phenomenological approach, the collection of data extracted
from a comparison to experiments, is not sufficient for assessing
the global behavior of $l_{{\rm ff}}$. Nevertheless, this indicates
the range of $l_{{\rm ff}}$ values for such system configurations
and gives a first hint as how it could behave as a function of the
AMF characteristics. Indeed, we mentioned earlier that $l_{{\rm ff}}$
may be inferred from experiments using the saturated temperature elevation
($\Delta T_{{\rm max}}$) and the corresponding SAR $\mathfrak{S}\left(T_{{\rm max}}\right)$.
This implies that in this long-time regime, $l_{{\rm ff}}$ should
follow the behavior of the SAR, in terms of its dependence on $h,\omega$
and the NM volume distribution ($D_{m},\delta$). In the intermediate
times, usually explored in experiments and reported in Figs. \ref{fig:ShahFit},
\ref{fig:MuraseFit}, the behavior of $l_{{\rm ff}}$ is subtler.
Indeed, as far as we can tell from these data for maghemite and magnetite,
$l_{{\rm ff}}$ seems to be a monotonously increasing function of
the AMF frequency while it exhibits a non regular variation with respect
to the amplitude. 

All in all, here we provide a first attempt to characterize this coefficient
but further experimental and theoretical investigations are required
in order to build a function for $l_{{\rm ff}}$ in terms of the characteristics
of both the AMF and the NM assembly.

\section{Conclusion}

We have built a simple formalism for studying the heat generated by
magnetic nanoparticles subjected to an alternating magnetic field
and its transfer to the sample's environment, with the assumption
that temperature equilibration within the ferrofluid is much faster
than the heat exchange between the ferrofluid and its external environment.
This formalism is based on i) the linear dynamic response of the (non-interacting)
assembly of nanomagnets given by the ac susceptibility which renders
the specific absorption rate of the system, and ii) a balance (phenomenological)
equation that describes the heat transfer at the interface between
the ferrofluid and its environment with the help of Newton's law of
cooling. The latter involves a heat transfer coefficient that helps
characterize the heat flow and the rate of temperature elevation within
the ferrofluid. The balance equation has been numerically solved and
a good approximate analytical solution has been provided.

Using our formalism, we studied the temporal profile of the temperature
rise as a function of the magnetic field frequency and amplitude and
have favorably compared it to the available experiments on magnetite
and maghemite ferrofluids. We have thus confirmed that an increase
of these parameters enhances the specific absorption rate and thereby
the temperature elevation in the prototypical ferrofluids studied
in experiments, namely maghemite and magnetite in water. We have also
discussed the behavior of the Newton heat-transfer coefficient for
these samples using the available data for various amplitudes and
frequencies of the magnetic field. The data collected by fitting our
theoretical results to the experimental curves provide a few hints
regarding the behavior of this coefficient but are not sufficient
to draw a general or universal tendency. For the latter, one should
also take into account the mechanisms of heat diffusion within the
nanofluid.

We suggest that this Newton's coefficient, introduced here to quantify
the heat transfer from the ferrofluid to its environment, could be
used as a new handle for optimizing the temperature rise in ferrofluids.
Its experimental studies should provide us with valuable hints for
building more refined models to better describe the mechanisms of
heat generation and its diffusion in ferrofluids used in applications
such as magnetic hyperthermia. In particular, its dependence on the
volume fraction of nanomagnets and their mutual interactions in the
ferrofluid is crucial for such applications.

\subsection*{Data availability }

The data that support the findings of this study are available from
the corresponding author upon reasonable request. \pagebreak{} 

\bibliographystyle{apsrev} \expandafter\ifx\csname natexlab\endcsname\relax
\global\long\def\natexlab#1{#1}%
\fi \expandafter\ifx\csname bibnamefont\endcsname\relax 
\global\long\def\bibnamefont#1{#1}%
\fi \expandafter\ifx\csname bibfnamefont\endcsname\relax 
\global\long\def\bibfnamefont#1{#1}%
\fi \expandafter\ifx\csname citenamefont\endcsname\relax 
\global\long\def\citenamefont#1{#1}%
\fi \expandafter\ifx\csname url\endcsname\relax 
\global\long\def\url#1{\texttt{#1}}%
\fi \expandafter\ifx\csname urlprefix\endcsname\relax
\global\long\def\urlprefix{URL }%
\fi \providecommand{\bibinfo}[2]{#2} \providecommand{\eprint}[2][]{\url{#2}}

\end{document}